\documentclass{article}
\usepackage[margin=0.5in]{geometry}
\usepackage{color}
\usepackage{amsmath,amssymb,graphicx}
\usepackage{authblk}
\usepackage{cite}
\usepackage[version=3]{mhchem}
\begin{document}
\title{Chiral Huygens metasurfaces \\for nonlinear structuring of linearly polarized light}

\author[1,2,*]{A. Cal\`a Lesina}
\author[1,2,3]{P. Berini}
\author[1,2,+]{L. Ramunno}
\affil[1]{Department of Physics, University of Ottawa, Canada} 
\affil[2]{Centre for Research in Photonics, University of Ottawa, Canada} 
\affil[3]{School of Electrical Engineering and Computer Science, University of Ottawa, Canada} 
\affil[*]{antonino.calalesina@uottawa.ca}
\affil[+]{lora.ramunno@uottawa.ca}
\maketitle
%
\begin{abstract}

We report on a chiral nanostructure, which we term a ``butterfly nanoantenna,'' that, when used in a metasurface, allows the direct conversion of a linearly polarized beam into a nonlinear optical far-field of arbitrary complexity. 
The butterfly nanoantenna exhibits field enhancement in its gap for every incident linear polarization, which can be exploited to drive nonlinear optical emitters within the gap, for the structuring of light within a frequency range not accessible by linear plasmonics. As the polarization, phase and amplitude of the field in the gap are highly controlled, nonlinear emitters within the gap behave as an idealized Huygens source. A general framework is thereby proposed wherein the butterfly nanoantennas can be arranged on a surface to produce a highly structured far-field nonlinear optical beam with high purity. A third harmonic Laguerre-Gauss beam carrying an optical orbital angular momentum of 41 is demonstrated as an example, through large-scale simulations on a high-performance computing platform of the full plasmonic metasurface with an area large enough to contain up to 3600 nanoantennas.

\end{abstract}

\section{Introduction}\label{sec1}

Classical optical lenses gradually change the properties of light, such as phase and polarization, during its propagation. This results in voluminous devices not suitable for photonic integrated circuits. Metasurfaces and flat optics aim to overcome this limit \cite{Yu2014,Meinzer2014}.
Abrupt changes in the properties of light can be introduced at the sub-wavelength scale by so-called meta-atoms, which are tiny scatterers or nanoantennas. 
By engineering each single emitter of a metasurface, complex structured beams can be created, such as vortex beams carrying orbital angular momentum (OAM) \cite{Allen1992,Padgett2011}. The interest in OAM of light is growing for potential applications in classical communications \cite{Wang2012}, quantum information processing and quantum cryptography \cite{Mair2001},  microscopy \cite{Furhapter2005}, laser machining \cite{Duocastella2012}, optical manipulation and particle trapping \cite{Grier2003}.
The creation of OAM states of light from a collection of meta-atoms requires an azimuthal phase tuning, which can be obtained by changing the geometry of the meta-atoms, {\it e.g.}, V-antenna with different apertures \cite{Yu2011} and subwavelength patterning \cite{Hakobyan2016}, or by using Archimede's spiral configurations \cite{Rui2015,Chen2015}. As an alternative, the Pancharatnam-Berry (geometric) phase can be exploited using a $q$ plate \cite{Marrucci2006} or a fixed meta-atom progressively rotated in a metasurface \cite{Karimi2014,Bouchard2014,Li2013,Ma2015}. In this case the incoming radiation has to be circularly polarized, and the spin angular momentum (SAM) is converted to OAM by spin-orbit coupling \cite{Bliokh2015,Ling2014}, including inside a laser cavity for the production of high purity OAM lasing modes \cite{Naidoo2015}.

Flat optics concepts have been recently applied to the nonlinear regime with the emerging interest in nonlinear metasurfaces and metamaterials \cite{Minovich2015}.
Nonlinear metasurfaces based on second harmonic generation (SHG) in split-ring resonators (SRR) have been recently demonstrated for beam shaping \cite{Keren-Zur2015}, and focusing and beam steering \cite{Tymchenko2015,Segal2015}.
Tuning the properties of the nonlinear emitters is fundamental for shaping the nonlinear beam phase front \cite{Li2015,Almeida2016}. 
In bulk materials, nonlinear susceptibilities can be predicted by the linear susceptibility, according to Miller's rule. In metamaterials and nanostructures the application of Miller's rule is not straightforward. For example, it holds for third harmonic generation (THG), but fails for SHG, which can be explained by the nonlinear scattering model \cite{OBrien2015}. Examples of nonlinear near-field control in nanostructures have recently been reported \cite{Ciraci2012,Wolf2016,Chen2015a,OBrien2015,Keren-Zur2015}. 
The nonlinear emission control is even more challenging in hybrid dielectric/plasmonic nanostructures, since the nonlinear optical generation can take place in the gap of the nanoantenna due to the presence of a nonlinear material, in the nonlinear dielectric surrounding the nanostructure, or in the nanostructure material itself \cite{Zayats2009}. 
The understanding of the nonlinear emission from hybrid nanostructures, is still under debate \cite{Utikal2011,Aouani2014,Metzger2014,Linnenbank2016}. Furthermore, the nonlinear emission in bare metal nanostructures strongly depends on the nanoantenna shape, preferring threadlike to bulky shapes \cite{Hentschel2012,Hanke2012}.


In this paper we focus on the nonlinear emission from the gap of a plasmonic nanoantenna which we call a butterfly nanonantenna. 
The butterfly nanoantenna exhibits uniform field enhancement in the gap for any incident linear polarization. 
This is due to the chirality of the nanoantenna, which exhibits field enhancement in the gap for only one circular polarization handedness.
The linear field in the gap is highly controllable, with an amplitude nearly independent of the angle $\theta$ of the incident linear polarization, and a phase linearly varying with $\theta$. 
This linear field can in turn be used to drive nonlinear optical processes, such as third harmonic generation (THG).
This results in an almost ideal Huygens source whose emission can also be highly controlled in amplitude, polarization and phase.
The unique properties of this butterfly nanoantenna ``meta-atom'' allow us to engineer the far-field of a generated nonlinear optical field, by designing the arrangement of thousands of such meta-atoms to create a chiral metasurface. Moreover, the nonlinear optical emission can be at frequencies that are not within the plasmonic bandwidth of the nanoantennas, allowing us to structure light at frequencies that would not otherwise be accessible via only the linear response of plasmonic devices.
In this way the nanoantenna is metallic in the linear regime, to exploit plasmonic field confinement \cite{Novotny2011,Gramotnev2010}, and dielectric in the nonlinear regime, to allow almost free propagation of the nonlinear fields into the far-field \cite{Yang2014,Arbabi2015}. 

We present a framework for the design of Huygens metasurfaces based on the concept of the Pancharatnam-Berry phase for the production of complex nonlinear beams carrying optical OAM at frequencies outside the plasmonic region.  Using a nonlinear optical process to generate the structured light results in a far-field beam of very high purity. The inherent chirality of the butterfly meta-atoms allows us to excite the metasurface with a linearly polarized incident wave at any angle; unlike previous schemes circularly polarized incident beams are not required.
The tight control of the near-field (linear and nonlinear) allows us to scale up to high order OAM states. 
Our demonstration is focused on Laguerre-Gauss beams but is valid also for Hermite-Gauss beams. 
A full numerical simulation is necessary because the structure cannot be simplified due to the lack of symmetries.
Through large scale simulations on a high-performance platform of thousands of gold butterfly antennas, we demonstrate the creation of a third harmonic beam with an OAM of 41.

\section{Butterfly nanoantenna}\label{sec2}

\begin{figure}[htbp]
\centering
\includegraphics[width=0.25\textwidth]{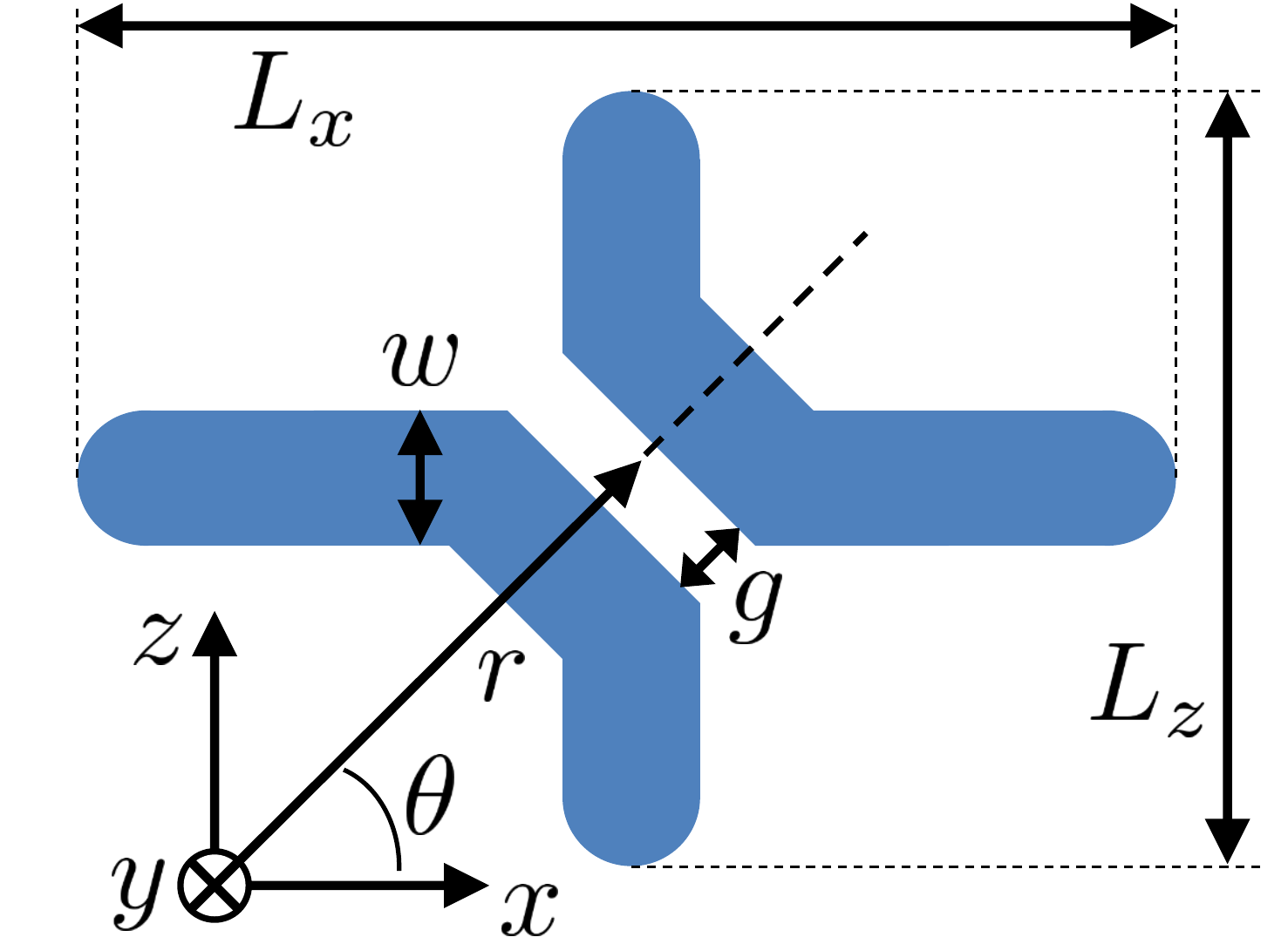}
\caption{Top view of the LH butterfly nanoantenna for gap a axis oriented at $\theta=45^{\circ}$.}
\label{fig1}
\end{figure}

The butterfly nanoantenna is sketched in front-view in Fig. \ref{fig1}. It is composed of bent metal strips of width $w$, thickness $t$, and rounded edges to minimize divergence effects in the electric field. The structure is asymmetric, {\it i.e.}, $L_x\neq L_z$, where $L_x$ and $L_z$ are the lengths of the structure along the $x$- and $z$-axis, respectively. The gap has size $g$ and the gap normal is oriented at $\theta$ with respect to the $x$-axis. 
 
 

\begin{figure}[htbp]
\centering
\includegraphics[width=0.8\textwidth]{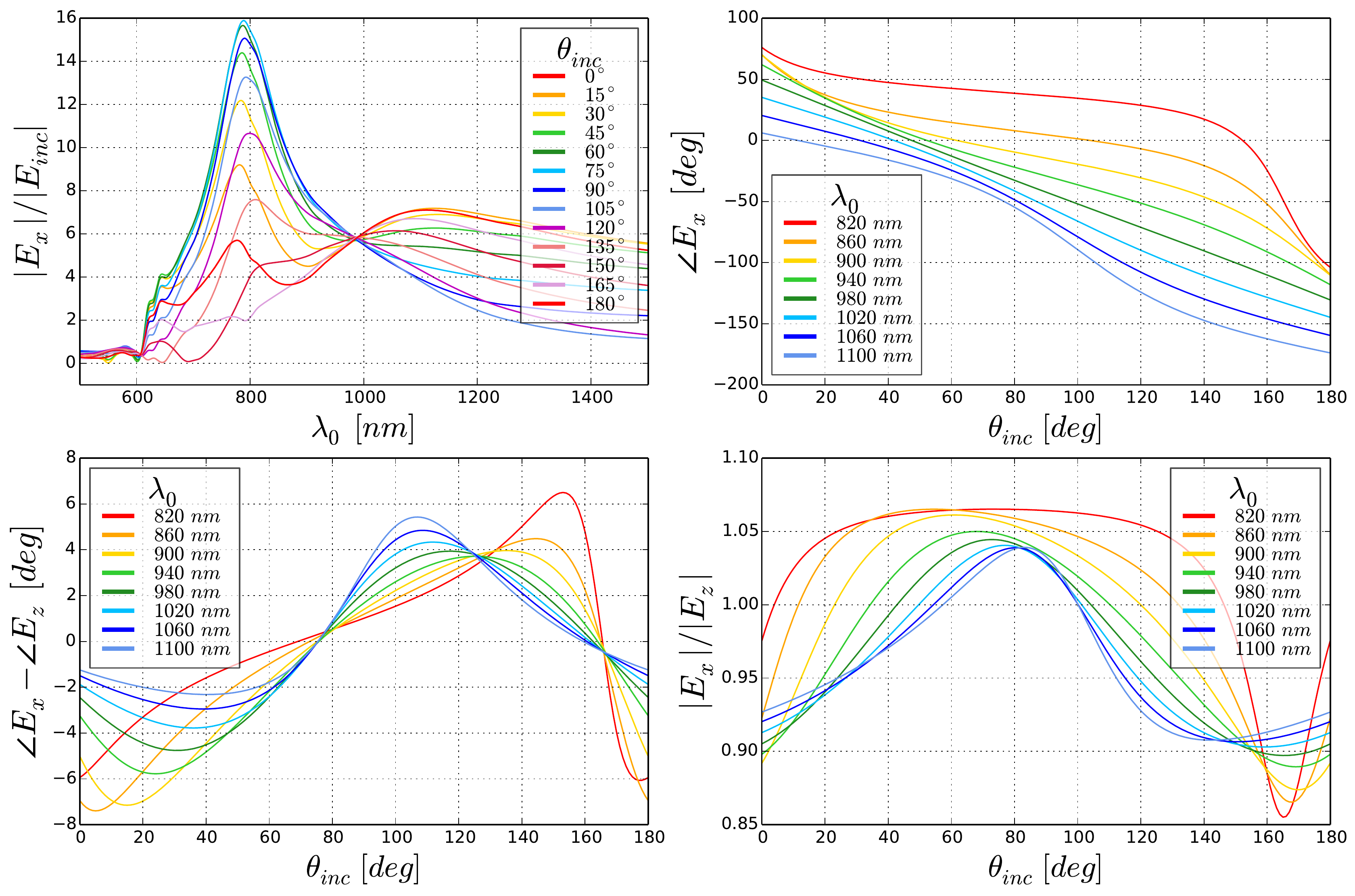}
\caption{Butterfly nanoantenna: (a) field enhancement $|E_x|/|E_{inc}|$ varying $\lambda_0$ and $\theta_{inc}$, (b) phase of $E_x$ varying $\theta_{inc}$ and $\lambda_0$, (c) phase difference between the components $E_x$ and $E_z$, (d) $|E_x|/|E_z|$ ratio.}
\begin{picture}(0,0)
\put(-205,195){(a)}
\put(5,195){(b)}
\put(-205,55){(c)}
\put(5,55){(d)}
\end{picture}
\label{fig2}
\end{figure}
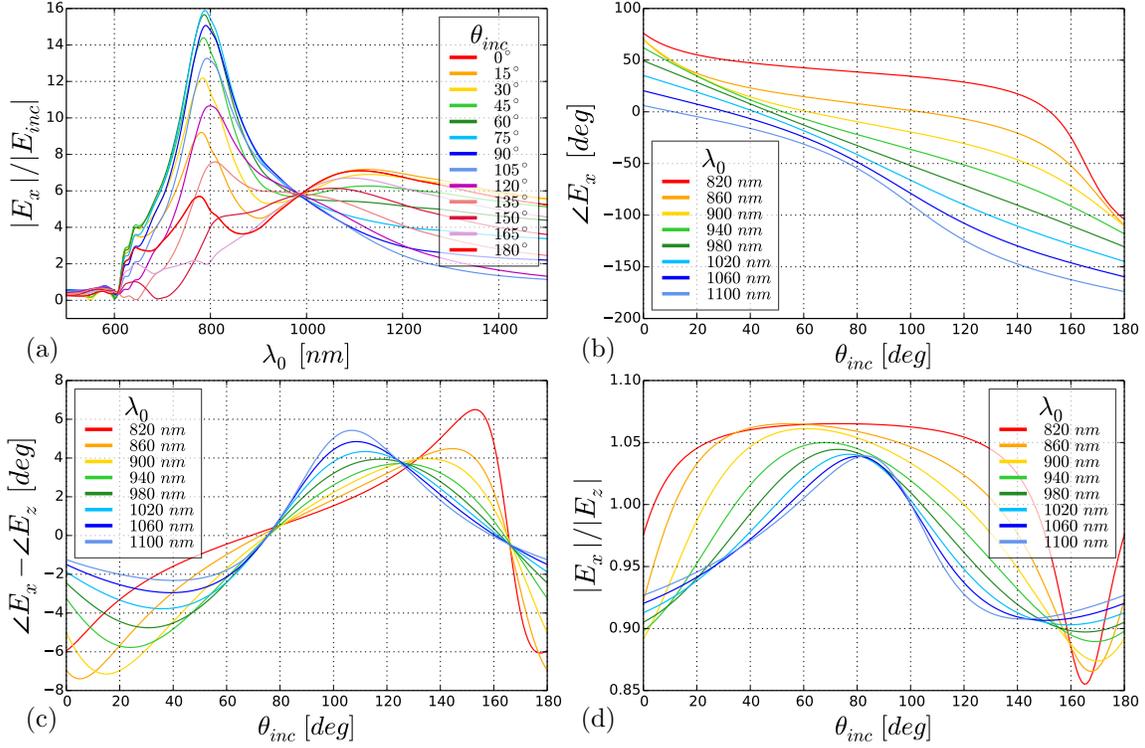

We consider a gold butterfly nanoantenna uniformly embedded in a generic bulk medium for demonstration purposes. This dielectric material has the dispersion of \ce{SiO2}, and relatively high $\chi^{(3)}_{diel}$ such that the nonlinear contribution from the gap predominates over the nonlinear contribution from gold. An example of a real material with these properties is \ce{ITO} which has a similar refractive index and can exhibit a large third order nonlinear response \cite{Alam2016}. 
We perform a broadband linear analysis to find the working wavelength of the nanoantenna. 
The two lengths $L_x$ and $L_z$ are responsible for two different resonances which create a crossing point hybridization mode (off-resonance). The crossing point can be seen in Fig. \ref{fig2}(a), which plots the $|E_x|$ enhancement in the gap with respect to the incident field $|E_{inc}|$ as a function of the free-space wavelength $\lambda_0$ for different linear polarization angles $\theta_{inc}$ of the incident wave. 
We sampled the field at one point in the middle of the gap; the field in the gap is nearly uniform because we work with the lowest order gap mode. A butterfly nanoantenna design that optimizes the crossing point has $L_x=300$ nm, $L_z=200$ nm, $w=60$ nm, $t=85$ nm, and $g=10$ nm. We found that a length ratio of $L_x/L_z=3/2$ 
guarantees the existence of the crossing point. For the parameters above, the crossing point occurs at the working wavelength of $\lambda_c=985.5$ nm ($\omega_c=304.2$ THz), and will be taken as the wavelength of the pump signal for the nonlinear generation process. 
At $\lambda_c$, the enhancement in the gap $|E_x|/|E_{inc}|$ is constant with varying $\theta_{inc}$. In Fig. \ref{fig2}(b) we report the phase of $E_x$ for different $\lambda_0$; we observe that at $\lambda_c$ the phase variation is linear with a slope of $-1$. This indicates the chirality of the nanoantenna, left-handed (LH) in this case; for right-hand (RH) chirality we would have a slope of $+1$. 
In Fig. \ref{fig2}(c) we plot the phase difference between the components $E_x$ and $E_z$, and observe an excursion that remains within $\sim 6^{\circ}$. In Fig. \ref{fig2}(d) we plot the ratio $|E_x|/|E_z|$ and observe that it remains close to 1. The $E_x$ and $E_z$ components thus have nearly the same amplitude and phase, which means the field in the gap is linearly polarized field in the gap directed at $\theta=45^{\circ}$. The direction of the field enhancement in the gap is constrained by the small gap to be parallel to the gap normal. 
The single nanoantenna was simulated with periodic boundary conditions (PBCs). 
The use of PBCs gives a good approximation to the case where the orientation of the meta-atom in the metasurface is slowly varying \cite{Byrnes2015}. The inter-distance $a=420$ nm used in these calculations is sub-wavelength; a lattice spacing of the order of the wavelength would create lattice resonant modes and interferences modifying the desired phase in the gap. 

\begin{figure}[htbp]
\centering
\includegraphics[width=0.24\textwidth]{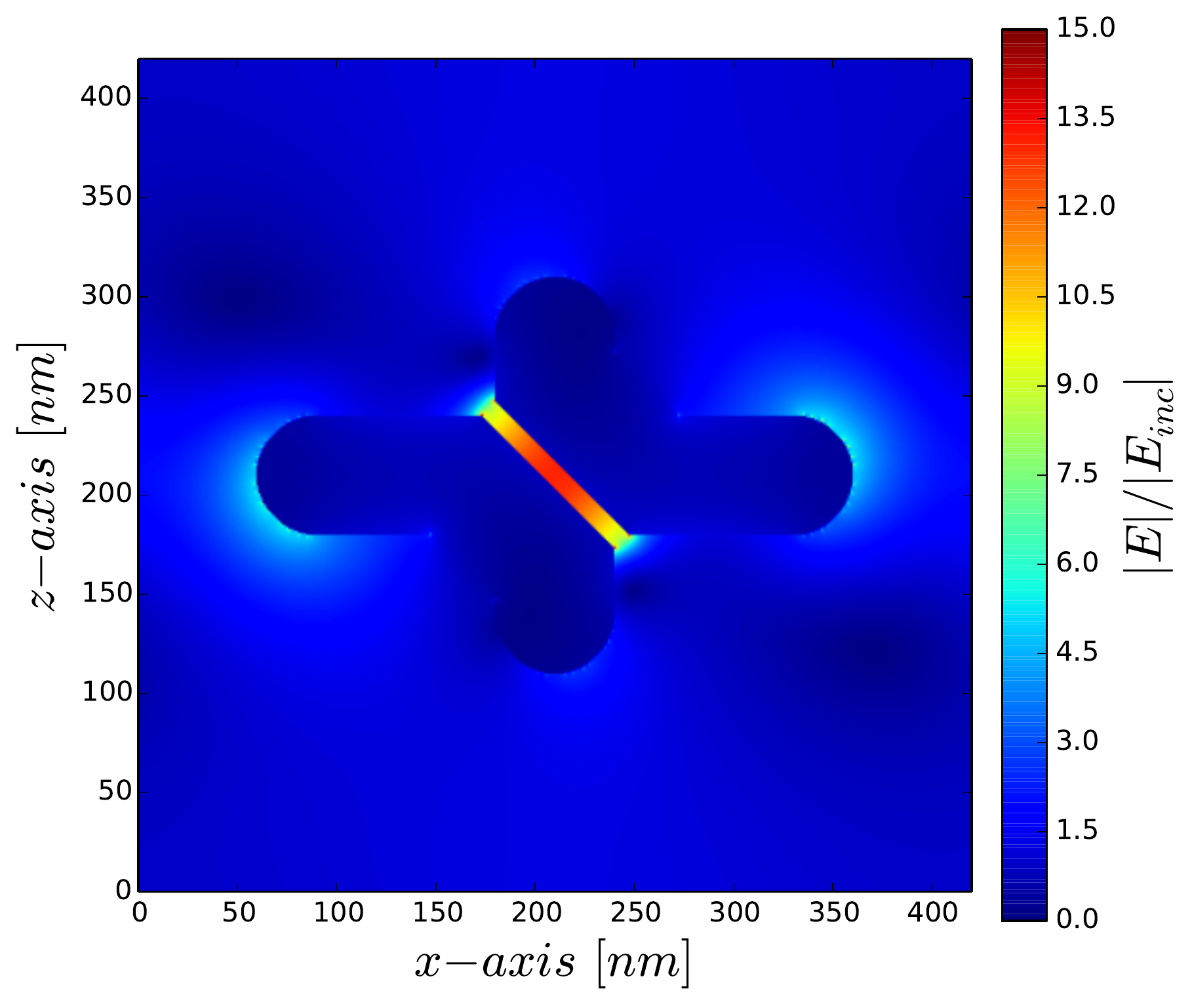}
\includegraphics[width=0.24\textwidth]{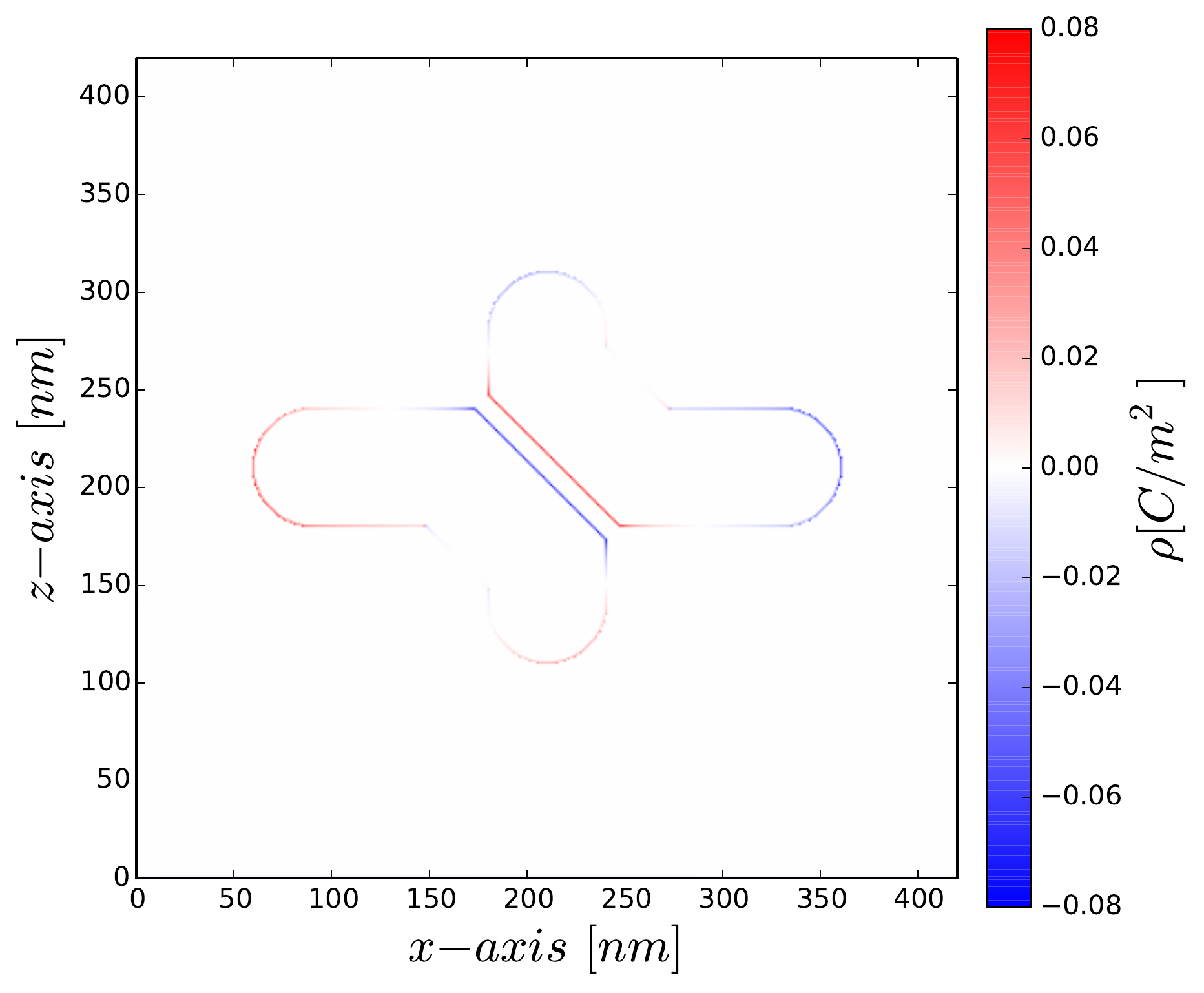}
\includegraphics[width=0.24\textwidth]{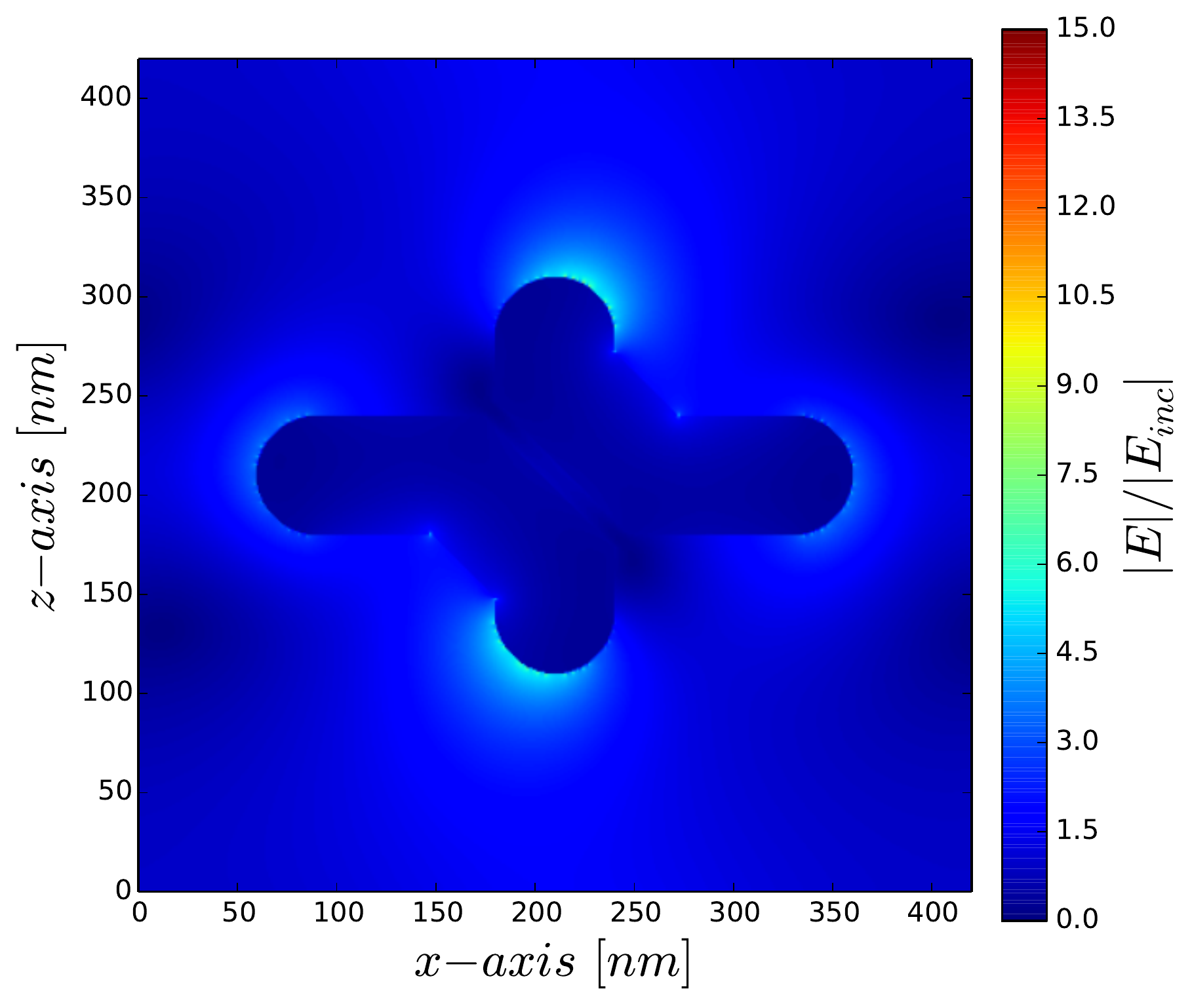}
\includegraphics[width=0.24\textwidth]{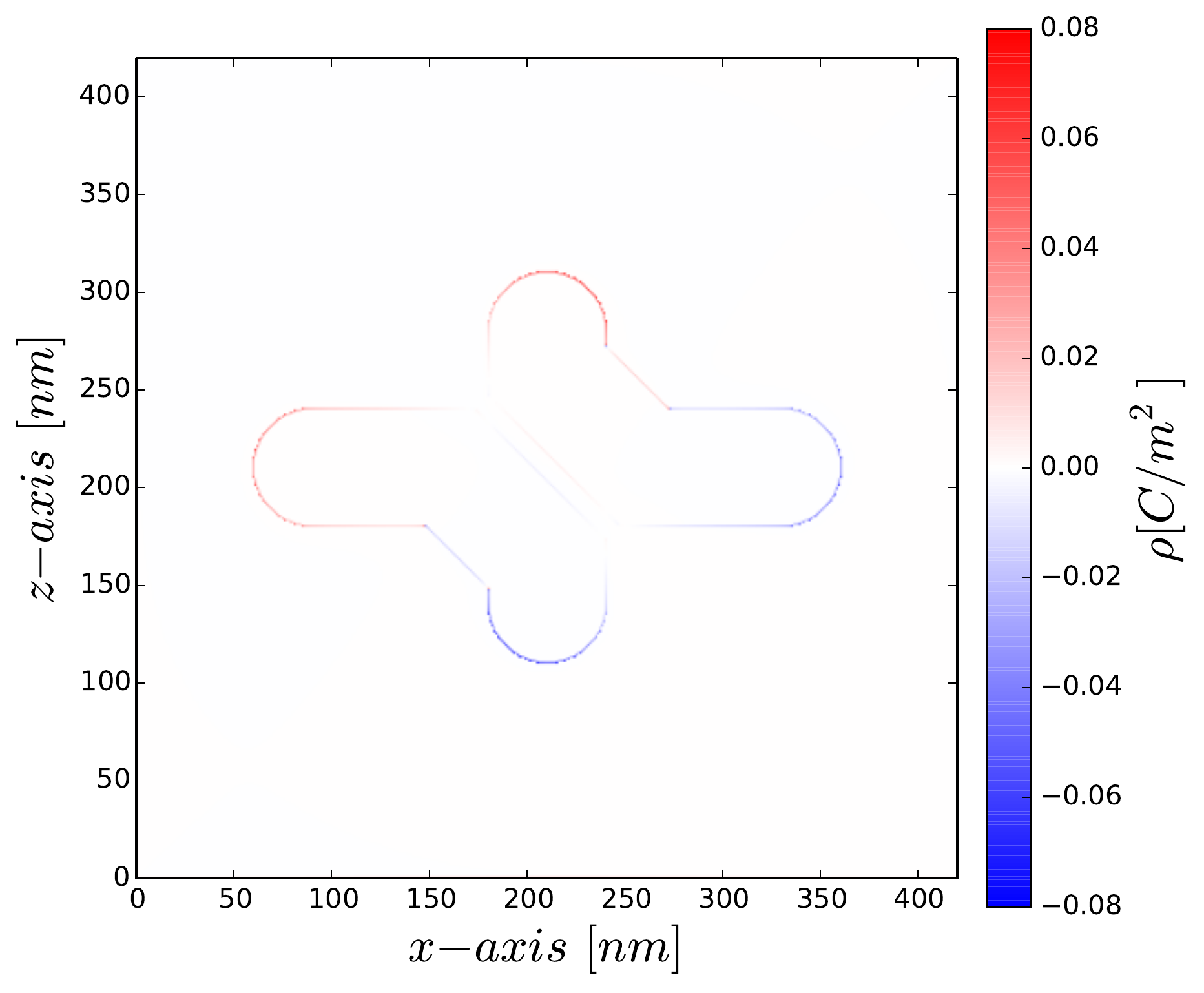}
\caption{Butterfly nanoantenna: (a) field enhancement $|E|/|E_{inc}|$ at $\omega_c$ for LCP excitation, (b) surface charge density for LCP excitation, (c) field enhancement $|E|/|E_{inc}|$ at $\omega_c$ for RCP excitation, (d) surface charge density for RCP excitation.}
\begin{picture}(0,0)
\put(-266,50){(a)}
\put(-133,50){(b)}
\put(0,50){(c)}
\put(135,50){(d)}
\end{picture}
\label{fig3}
\end{figure}
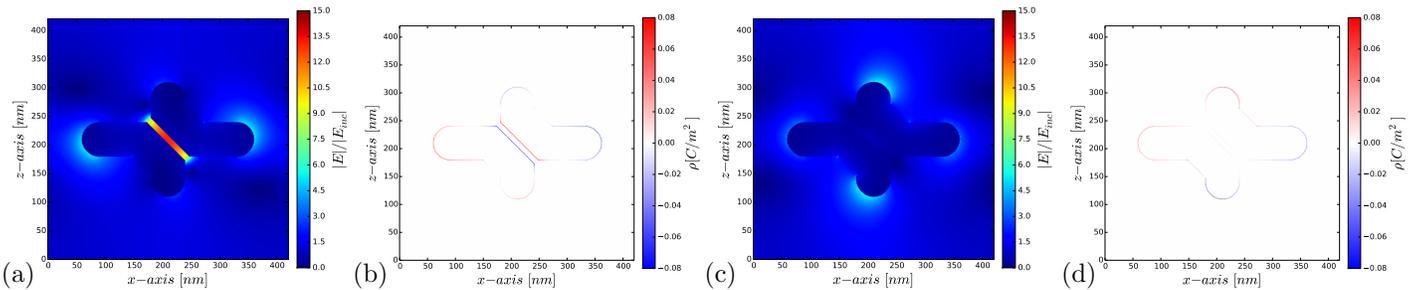

The LH butterfly ($L_x>L_z$) exhibits optimal coupling to left circular polarization (LCP) in terms of producing field enhancement in the gap; the LH butterfly will be used throughout the paper.
Exciting an LH butterfly with right circular polarization (RCP) produces negligible field enhancement in the gap. In Figs. \ref{fig3}(a,c) we show the field enhancement at $\omega_c$ for LCP and RCP, respectively. The field enhancement in the gap under LCP illumination is one order of magnitude higher than under RCP excitation. In Fig. \ref{fig3}(b,d) we give the surface charge density showing a strong dipole oscillation in the gap for LCP excitation. 
In \cite{CalaLesina2015} we considered a similar unit cell to build a metasurface for difference frequency generation, which for simplicity was assumed square ($L_x=L_z$). The mirror symmetry with respect to the $\theta=135^{\circ}$ axis in that case prevented the production of a field enhancement in the gap for $135^{\circ}$ incident linear polarization. 
Breaking the symmetry of the structure by using $L_x\neq L_z$ is fundamental to producing a field enhancement in the gap for every incident linear polarization. 

The nonlinear analysis is performed by LCP continuous wave (CW) excitation at $\omega_c$ and generating a THG signal at $3\omega_c$ by the instantaneous isotropic Kerr effect.
The CW excitation allows us to isolate the THG component specifically at $3\omega_c$, avoiding frequency mixing and dispersive effects in the third order susceptibility.
Simulations of hybrid dielectric/plasmonic nanoantennas show that $\chi^{(3)}_{diel}\geq10^{-1}\cdot\chi^{(3)}_{Au}$ is a sufficient condition for neglecting the nonlinearity in gold. 
The nonlinear field inherits the polarization of the linear field, and its phase is $\angle E_x(3\omega)= 3\angle E_x(\omega)$. The third harmonic falls in the ultraviolet range where plasmonic effects are not present in gold, which behaves rather as an almost transparent dielectric. This makes the THG hot-spot in the gap of the butterfly nanoantenna well-approximated as a Huygens source.
The nonlinear dipole created in the gap of the butterfly nanoantenna thus can be used as the building element of a metasurface to produce complex nonlinear beams, {\it e.g.}, carrying orbital angular momentum, which we now demonstrate in the next section. 


\section{Nonlinear metasurfaces}\label{sec3}


A beam can be envisioned as the far-field radiation produced by a distribution of Huygens sources, {\it i.e.}, radiating dipoles. Tuning the amplitude, polarization and phase of the dipoles in the near-field, can produce a desired structured beam in the far-field.
Before considering metasurfaces with butterfly antennas, we first introduce a formalism to describe the distribution of idealized radiating dipoles, {\it e.g.}, point source apertures with an assigned electric field (or alternatively, current density sources).
We use Cartesian $(x,z)$ and cylindrical coordinates $(r,\phi)$ in the $xz$ plane, with $\vec{r}=x\hat{x}+z\hat{z}$ and $\phi=tan^{-1}(z/x)$. 
We consider a circular array of dipoles, with the origin located at the centre of the array. The dipoles are arranged in the $xz$ plane following a square lattice of lattice constant $a=420$ nm as determined above. The dipoles are distributed starting from the positive $x$-axis following the phasor equations
\begin{equation}
\begin{split}
\\E_x(\vec{r})=|E|\cos(\alpha+\gamma\phi)e^{in\gamma\phi},
\\E_z(\vec{r})=|E|\underbrace{\sin(\alpha+\gamma\phi)}_{polarization}\underbrace{e^{in\gamma\phi}}_{phase},
\end{split}
\label{eq1}
\end{equation}
for $\vec{r}=a(n_x\hat{x}+n_z\hat{z})$, where $n_x$ and $n_z$ are integers in $\{-N_d,...,N_d\}$, $n_x^2+n_z^2\leq N_d^2$, $N_d$ is the number of dipoles along the radius, $\phi=tan^{-1}(n_z/n_x)$, $|E|$ is the electric field amplitude associated with the dipole, $\alpha$ is the orientation of the dipoles along the positive $x$-axis, $\gamma$ is the number of full rotations of the dipole polarization per $2\pi$, $n$ is the order of the nonlinearity, and $i=\sqrt{-1}$. 
 
\begin{figure}[htbp]
\centering
\includegraphics[width=0.35\textwidth]{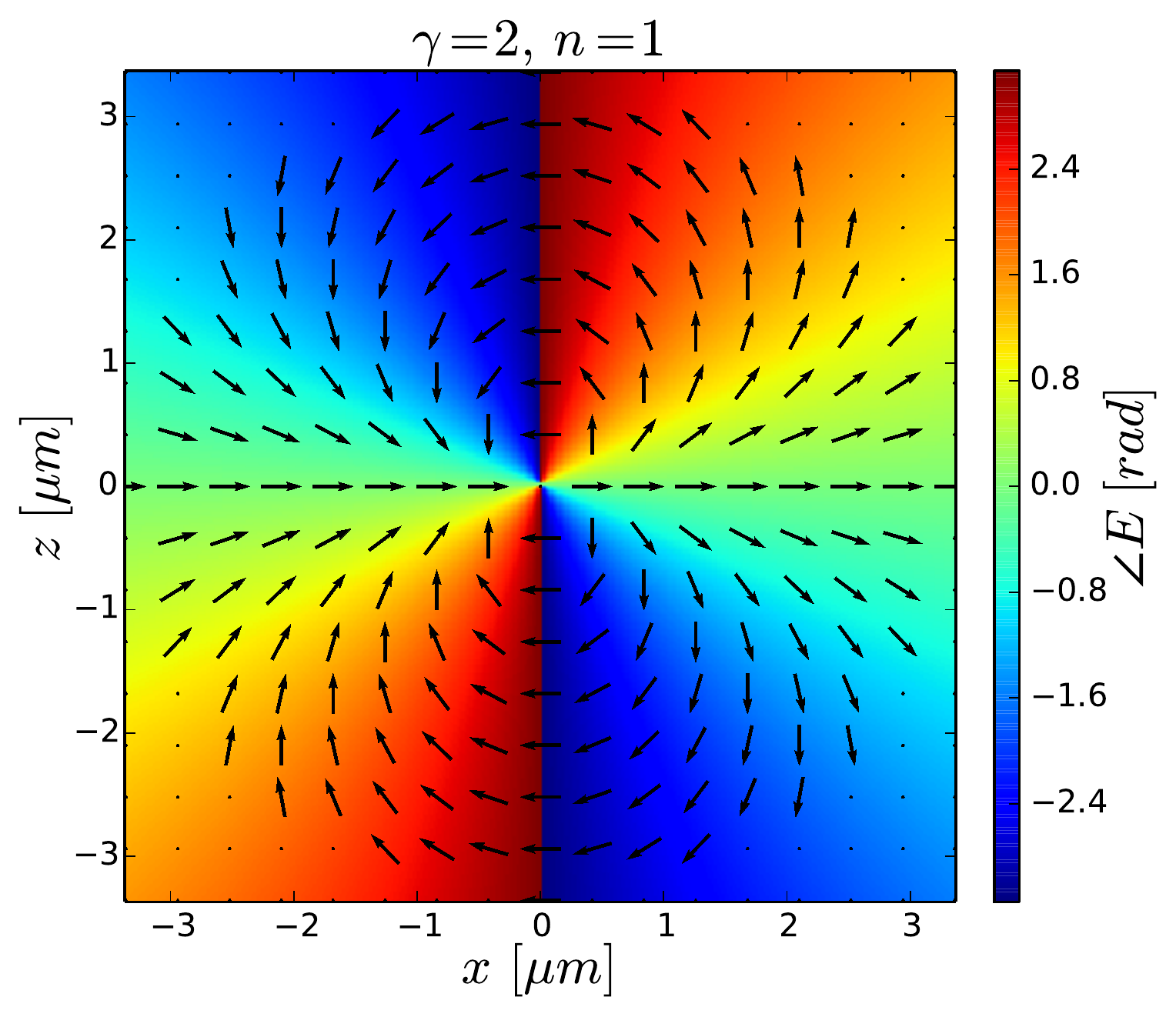}
\includegraphics[width=0.35\textwidth]{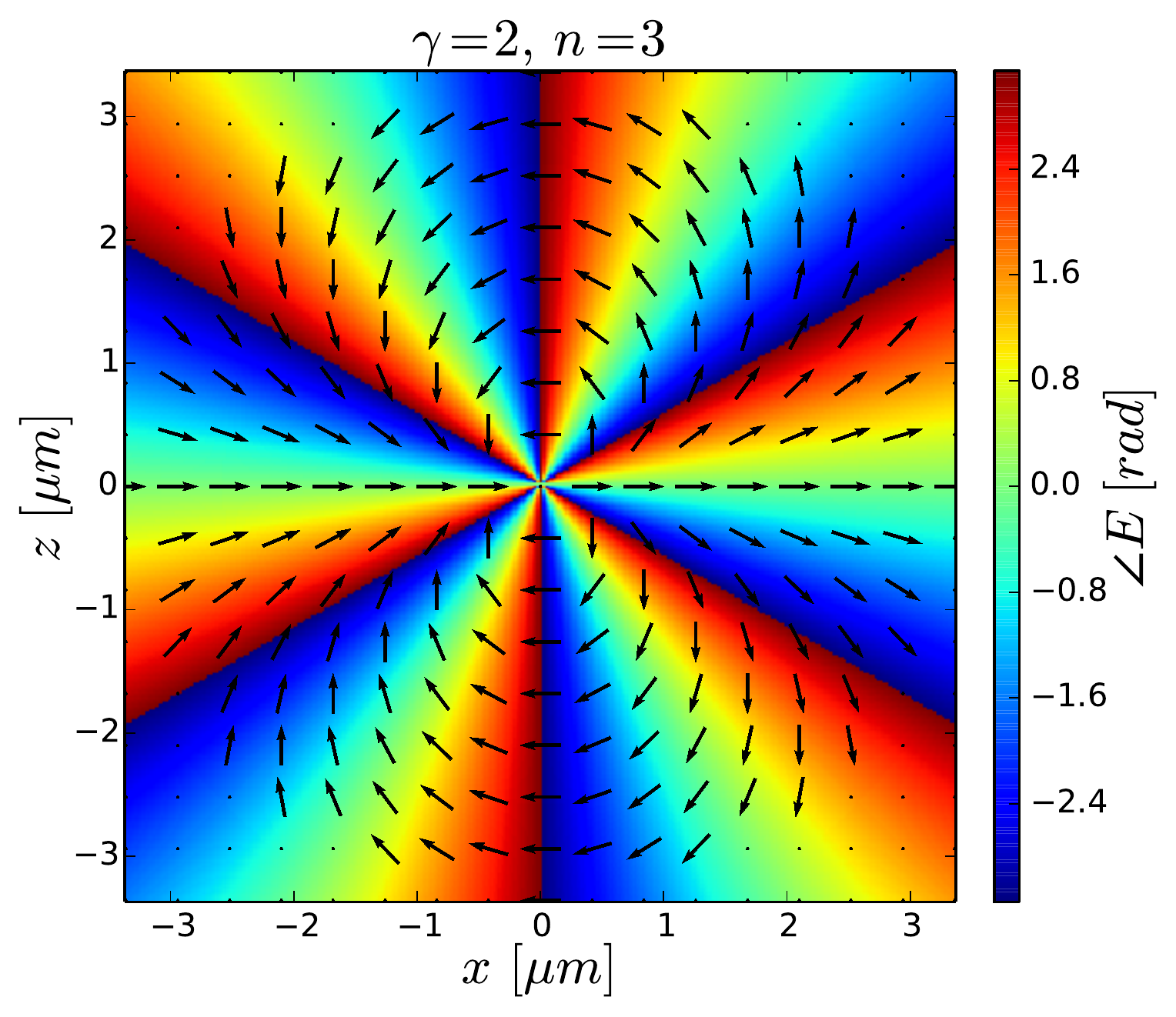}
\includegraphics[width=0.285\textwidth]{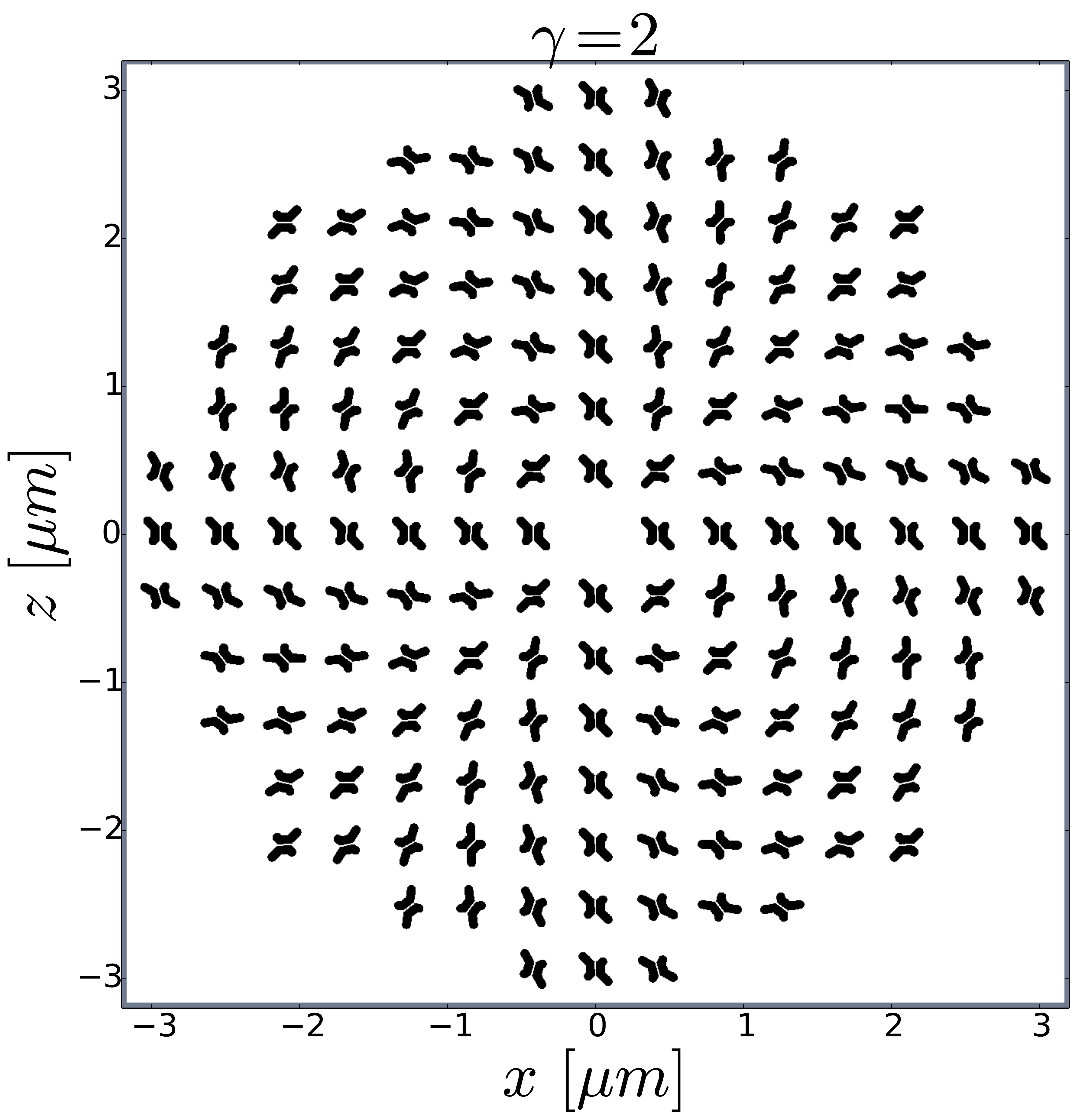}
\caption{$\gamma=2$: (a) linear ($n=1$) array, (b) nonlinear ($n=3$) array, (c) butterfly metasurface layout.}
\begin{picture}(0,0)
\put(-260,35){(a)}
\put(-65,35){(b)}
\put(120,35){(c)}
\end{picture}
\label{fig4}
\end{figure}
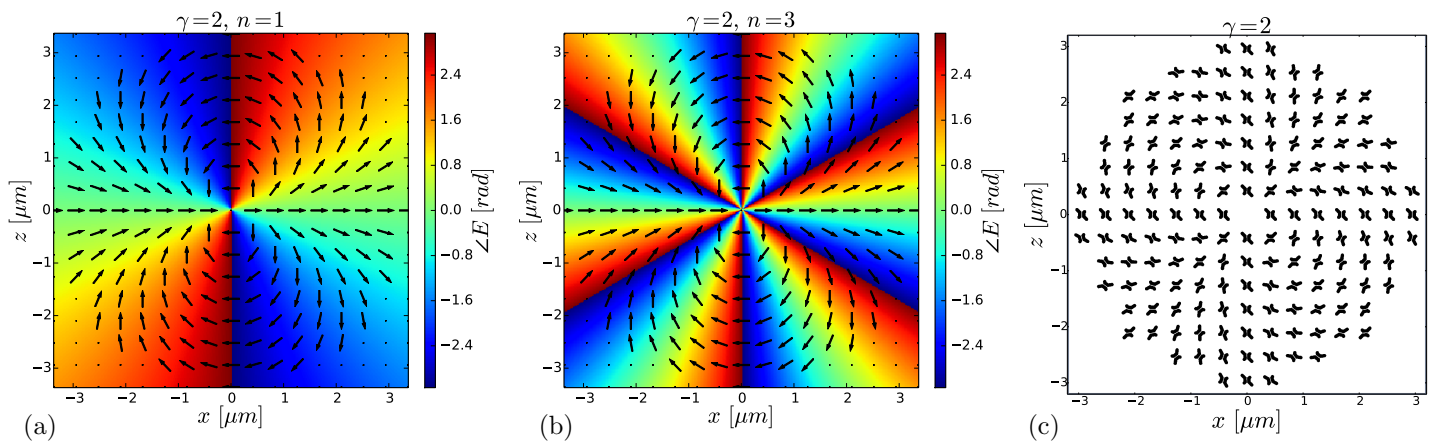

In Figs. \ref{fig4}(a,b), dipole sources are depicted as arrows, the orientation of the arrow represents its polarization and the underlying color its phase. 
We consider $\alpha=0$ and a constant amplitude $|E|$. The polarization of a single dipole is directed at angle $\alpha+\gamma\phi$ and its phase is $n\gamma\phi$. We can distinguish two topological charges: $\gamma$ for the polarization of the dipole and $n\gamma$ for the phase.
Linear and nonlinear dipoles have the same polarization, but they differ in phase and amplitude.
For illustrative purposes, we show distributions of dipoles for the linear and nonlinear cases in Figs. \ref{fig4}(a,b), respectively, for $\gamma=2$ and $N_d=8$. 
The Huygens sources in the arrays of Figs. \ref{fig4}(a,b) can be produced with butterfly nanoantennas when they are positioned such that the gap normal is oriented along the dipole polarization, {\it i.e}, $\theta=\alpha+\gamma\beta$.
Through this substitution, we obtain the butterfly metasurface of Fig. \ref{fig4}(c).


In a real metasurface $|E|$ depends on the location of the butterfly antenna relative to the illuminating beam and on the gap size $g$, which is taken as constant throughout the paper. A plane wave excitation produces $|E|$ $\sim$ constant in the gap as predicted from Fig. \ref{fig2}(a). This is a good approximation to the case of illumination by a loosely focused Gaussian beam.
We consider the butterflies to be embedded within a homogeneous medium for simplicity. Embedding the metasurface into the surface of the nonlinear material and having air on the other side, blue-shifts the crossing point to $\lambda_0=915$ nm due to the lower effective refractive index. We excite the structure by a LCP CW signal at $\omega_c$ propagating along $y$ producing a nonlinear beam at $3\omega_c$ in the forward direction. 
The butterfly metasurface can generate structured nonlinear beams for both LCP and RCP excitations. Due to the chirality of the metasurface, the nonlinear far-field generated with RCP illumination is one order of magnitude less intense. The structured beam generated under LCP excitation is due to nonlinear dipoles localized in the gaps. For RCP excitation the contributions mainly come from field enhancement outside the gap. 


In general, with the arrangement described by Eq. \ref{eq1}, Laguerre-Gauss ($LG$) modes are obtained. $LG_{lp}$ modes describe the state of single photons by the quantum numbers $\sigma=\pm 1$, $l=0,\pm1,\pm 2,...$ and $p=0,1,2,...$ associated with SAM, OAM, and radial eigenstates, respectively \cite{Harris2015}. 
Hermite-Gaussian modes can be obtained by using square/rectangular distributions of meta-atoms instead of radial/circular ones.

Working in cylindrical coordinates, we found that the OAM state of the innermost intensity ring, figuring as a phase term $e^{il\phi}$ in the far-field beam, is given by

\begin{equation}
l=\sigma(1-\gamma+n\gamma),	
\end{equation}
where $\sigma=\pm 1$ for incident LCP or RCP, respectively.
The OAM order can be increased in two ways: increasing $\gamma$ or increasing $n$.
Since we are considering $n=3$, we investigate a specific example where we reach high order OAM by increasing $\gamma$. We consider $\gamma=20$ and $N_d=30$ (Fig. S. 1 -- supplementary section), resulting in $l=41$.
When $\gamma$ is large the butterfly nanoantennas in the centre of the metasurface cannot resolve the topological charge. The far-field beam can be polished by removing the innermost butterflies that do not properly resolve the desired number of rotations, that is, by removing the nanoantennas inside the circle of radius $r=a\sqrt{n_x^2+n_z^2}$, such that $\frac{2r\pi}{\bar{a}}>2\gamma$ (Nyquist condition), where $\bar{a}\sim a$ is the average distance between nanoantennas along the circle.

\begin{figure}[htbp]
\centering
\includegraphics[width=1\textwidth]{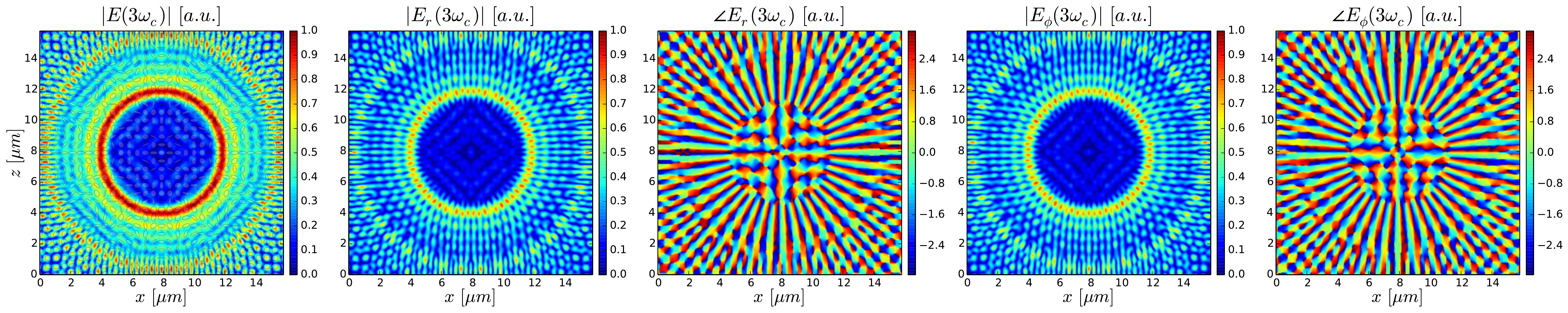}
\includegraphics[width=1\textwidth]{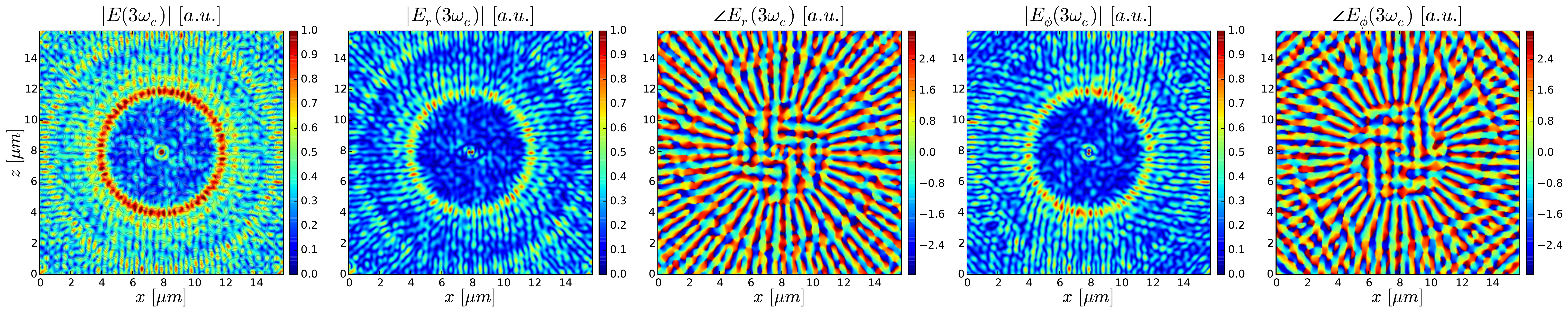}
\caption{$\gamma=20$, nonlinear far-field beam ($l=41$): analytical arrangement of dipoles (top) versus numerical modelling (FDTD) of the corresponding arrangement of butterfly nanoantennas (bottom).}
\label{fig5}
\end{figure}

In Fig. \ref{fig5} top row, we show the far-field from an idealized set of nonlinear dipoles, where the far-field was calculated by a near-to-far transformation of the surface field distribution described by Eq. \ref{eq1}. 
The bottom row shows the nonlinear far field generated via FDTD simulations of the corresponding arrangement of butterfly nanoantennas embedded in third order nonlinear material.
The near-field distribution at $3\omega_c$ on a $xz$ plane cut was used to numerically calculate the far-field by near-to-far transformation. Movie 1 (Fig. S. 2 -- supplementary section) shows the time-domain evolution of an LCP plane wave interacting with
the metasurfaces for $\gamma = 20$, in a $xz$ plane cut through the middle of the gaps.
We observe that both methods show an OAM state of $l=41$. There is very good agreement between the two rows, indicating that the butterflies, even when arranged on the surface, do act as effective idealized Hugyens sources. The agreement is expected to improve by locally optimizing each butterfly on the metasurface. 
In Fig. \ref{fig5} we observe that the external far field intensity ring carries $l_2=79$. In general, the OAM state of the external ring is $l_2=l+2\sigma(\gamma-1)$. This higher order OAM state is due to the diffraction of a uniform plane wave by the metasurface.

\section{Conclusion}
We proposed a plasmonic nanoantenna to control nonlinear optical emission by the linear field enhancement in the gap. We used the nanoantenna as meta-atom in Huygens metasurfaces, and we demonstrated its applicability to arbitrarily structure the far-field radiation. Due to the chirality of the nanoantenna, only one circular polarization handedness enables the field enhancement in the gap and the consequent nonlinear emission. This results in the direct conversion of a linearly polarized wave to a nonlinear beam with arbitrary complexity.   
A highly pure Laguerre-Gauss beam carrying an orbital angular momentum of 41 at the third-harmonic of the linear exciting field was demonstrated. The beam synthesis framework is general and opens the door to applications requiring very high purity, and wavelengths not accessibly by linear plasmonics. In addition to the low order nonlinear processes investigated here, higher order processes may be considered, such as high harmonic generation from solids, which is very topical and could lead to the structuring of extreme ultraviolet light. 


\section{Methods}
The 3D full-wave simulations are performed by the finite-difference time-domain (FDTD) method exploiting high-performance computing (HPC).
The simulations are performed using an in-house parallel 3D-FDTD code \cite{CalaLesina2015a}.
We model the dispersion of gold by the Drude model plus two critical points (Drude+2CP), implemented in FDTD by the auxiliary differential equation (ADE) technique. We use a per-component sub-cell mesh, with a uniform grid of space step $\Delta x= \Delta y=\Delta z= 2$ nm. Linear broadband analyses are performed on the single butterfly nanoantenna by plane-wave pulse excitation and in-line discrete Fourier transform (DFT). The plane-wave is introduced by means of the total-field/scattered-field (TF/SF) method. The simulation domain is terminated by convolutional perfectly matched layers (CPML) absorbing boundary conditions and periodic boundary conditions (PBCs) are applied. 
The nonlinear monochromatic analysis of the full metasurface is performed by LCP excitation, using a CW signal at $\omega_c$ modulated by a Gaussian pulse.
The propagation is along the $y$ direction.
The metasurface is uniformly embedded in \ce{SiO2}, the linear dispersion is modeled by the Lorentz model using the ADE technique. The THG calculations run concurrently in the FDTD code introducing in the ADE framework of the Lorentz model \cite{Varin2016} the instantaneous Kerr nonlinear polarization terms 
\begin{equation}
\begin{split}
P_x(t) = \chi^{(3)}_{xxxx}E_x(t)|E(t)|^2,\\
P_y(t) = \chi^{(3)}_{yyyy}E_y(t)|E(t)|^2,\\
P_z(t) = \chi^{(3)}_{zzzz}E_z(t)|E(t)|^2,\\
\end{split}
\end{equation}
where $|E(t)|^2=E_x^2(t)+E_y^2(t)+E_z^2(t)$, we assume isotropic dispersionless nonlinearity, considering the third order susceptibility $\chi^{(3)}_{diel}=\chi^{(3)}_{xxxx}=\chi^{(3)}_{yyyy}=\chi^{(3)}_{zzzz}$, with $\chi^{(3)}_{xxxx}=\chi^{(3)}_{xxzz}+\chi^{(3)}_{xzxz}+\chi^{(3)}_{xzzx}$. We consider a fictitious nonlinearity in the dielectric such that $\chi^{(3)}_{diel}>10^{-2}\cdot\chi^{(3)}_{Au}$, and we neglect the nonlinearity in gold. 
The analytical far-field is obtained by near-to-far transformation (Fraunhofer diffraction or Fourier optics) of the near-field dipole array electric field distribution.
The numerical far-field is obtained by near-to-far transformation of the near-field distribution calculated by FDTD simulation. 
The $xz$ cut plane was taken $10$ nm after the metasurface to enclose the contributions from the entire gap region; similar results are obtained considering a plane cut through the middle of the gaps due to the uniformity of the field in the gap.
The near-to-far transformation is implemented in Cartesian coordinates by
\begin{equation}
\vec{E}(X,Z)=\sum_{N_x}\sum_{N_z}\vec{E}(x,z)e^{i\frac{2\pi}{\lambda_0}\frac{1}{R}(Xx+Zz)}\Delta x\Delta z,
\end{equation}
where $N_x$ and $N_z$ are the number of cells in the near-field along $x$ and $z$, respectively, $X$ and $Z$ are the coordinates in the far-field, and $R$ is the distance of the far-field plane; in our simulations we used $R=30$ $\mu m$.
The Cartesian coordinate field is then converted to cylindrical coordinates to visualize the OAM state. 
The near-to-far transformation runs on a GPU cluster exploiting OpenACC.
The largest simulation for $\gamma=20$ required $\sim 30\cdot 10^9$ FDTD Yee cells and $32k$ computing cores.
The simulations were performed on the IBM BlueGene/Q supercomputer in the Southern Ontario Smart Computing Innovation Platform (SOSCIP).
\appendix
\subsection*{Acknowledgements}
We acknowledge the Canada Research Chairs program, the Southern Ontario Smart Computing Innovation Platform (SOSCIP), IBM Canada Research and Development Centre, the Natural Sciences and Engineering Council of Canada (NSERC), and SciNet.
\subsection*{Authors contributions}
A.C.L. conceived the butterfly nanoantenna, designed and conducted the simulations, discussed results, and co-wrote the manuscript.
P.B. discussed results, and co-wrote the manuscript.
L.R. proposed the study, discussed results, and co-wrote the manuscript.

\subsection*{Competing financial interests}
The authors declare that they have no competing financial interests.

\bibliography{Butterfly_paper}{}
\end{document}